%% file: fnlim_atmp.tex
\documentclass[11pt,twoside]{./atmp}

\usepackage{amsmath,amssymb,comment,latexsym}
\usepackage{finalT}
\usepackage[all]{xy}
\usepackage{color}

\input fnlim.def

\copyrightnotice{0}{0}{0}{0} 

\begin{document}

\title[The fast Newtonian limit for perfect fluids]
{The fast Newtonian limit for perfect fluids}

\arxurl{0908.4455}

\author[T.A Oliynyk]{Todd A. Oliynyk}

\address{School of Mathematical Sciences\\
Monash University, VIC 3800\\
Australia}
\addressemail{todd.oliynyk@monash.edu.au}

\begin{abstract}
We prove the existence of a large class of dynamical solutions to the Einstein-Euler equations
for which the fluid density and spatial three-velocity converge to a solution of the Poisson-Euler
equations of Newtonian gravity. The results presented here generalize those of \cite{Oli06} to
allow for a larger class of initial data. As in \cite{Oli06}, the proof is based on a non-local
symmetric hyperbolic formulation of the Einstein-Euler equations which contain a singular
parameter $\ep=v_T/c$ with $v_T$ a characteristic speed associated to the fluid and $c$ the speed of light. Energy
and dispersive estimates on weighted Sobolev spaces are the main technical tools used to analyze the solutions
in the singular limit $\ep\searrow 0$.
\end{abstract}

\maketitle

\input intro.tex

\input red.tex
\input eul.tex
\input uniform.tex

\input limiteqns.tex
\input nlim.tex

\bibliographystyle{my-h-elsevier}

\end{document}

%% file: intro.tex
\sect{intro}{Introduction}

The Einstein-Euler equations, which govern a gravitating
perfect fluid, are given by
\eqn{EEeqnA}{
G^{ij} = \frac{8\pi G}{c^4} T^{ij}\,
\AND \nabla_{i} T^{ij} = 0,} where \eqn{EEdefsA}{ T^{ij} = (\rho +
c^{-2} p)v^i v^j + p g^{ij},  } with $\rho$ the fluid density,
$p$ the fluid pressure, $v^i$ the fluid four-velocity normalized
by $v^i v_i = -c^2$, $c$ the speed of light, and $G$ the
Newtonian gravitational constant. 
Defining
\eqn{epdef}{\ep = \frac{v_T}{c},} where $v_T$ is a typical speed associated with
the fluid, 
the Einstein-Euler \cutpage
\noindent equations, upon suitable rescaling \cite{Oli06}, can be written
in the form \leqn{EEeqn}{ G^{ij} = 2\ep^4 T^{ij} \AND
\nabla_{i} T^{ij} = 0, } where \eqn{EEdefs}{ T^{ij} = (\rho + \ep^2 p)v^i v^j
+ p g^{ij} \AND v^i v_i = -\frac{1}{\ep^2}. }
In this formulation, the fluid four-velocity
$v^i$, the fluid density $\rho$, the fluid pressure $p$, the
metric $g_{ij}$, and the coordinates $(x^i)$ $i=1,\ldots,4$ are
dimensionless.  By assumption, the $(x^i)$ are global Cartesian
coordinates on spacetime $M \cong \Rbb^3\times [0,T)$, where the $(x^I)$
$(I=1,2,3)$ are spatial coordinates that cover $ \Rbb^3$, and
$t=x^4/v_T$ is a \emph{Newtonian time coordinate} that covers the interval
$[0,T)$.
By a choice of units, we set
$v_T=1$.

The Newtonian limit for the Einstein-Euler system refers
to the limit of solutions of the Einstein-Euler systems in the limit
 $\ep \searrow 0$. In this limit, one expects that under reasonable assumptions
 solutions of the Einstein-Euler system should converge to
 a solution of the Poisson-Euler equations of Newtonian gravity:
 \lalign{newtPE}{
\del_t \rhot + \del_I(\rhot\wt^I) & = 0 && (\del_I := \del_{x^I}) , \label{newtPE.1}\\
\rhot(\del_t \wt^J + \wt^I\del_I \wt^J) & =
-(\rhot\del^J\Phit + \del^J\pt)  && (\del^J := \delta^{JI}\del_I ),\label{newtPE.2} \\
\Delta \Phit &=    \rhot && ( \Delta := \delta^{IJ}\del_I\del_J), \label{newtPE.3}}
where $\rhot$, $\pt$, and $\wt^J$ are the fluid density, pressure,
and three-velocity, respectively.

The difficulty of analyzing the Newtonian limit arises
from the fact that the limit $\ep \searrow 0$ is singular.
The first general rigorous result on the Newtonian limit
without any symmetry assumptions is \cite{Ren94}.  There, it is shown that there
exists a wide class of solutions to the Einstein equations
coupled to Vlasov matter that have a well defined Newtonian
limit as $\ep\searrow 0$. This work is based on
an elliptic-hyperbolic formulation of the Einstein-Vlasov
equations in a maximal slicing gauge. In \cite{Oli06},
we used a different approach
to prove existence of a large class of non-stationary
solutions to the Einstein-Euler equations which have a
Newtonian limit.

The main aim of this article is to establish the existence of a Newtonian
limit for solutions to the Einstein-Euler equations under weaker conditions on the
initial data as compared to either \cite{Oli06} or \cite{Ren94}. The purpose
for this is twofold. First, it is of both theoretical and practical
interest to understand the most general situations possible for which
Newtonian gravity provides an acceptable approximation to full
Einstein theory. Second, the techniques developed here can
be used to improve the results of \cite{Oli08} on the existence
of post-Newtonian expansions. In \cite{Oli08}, it was shown
that there exists a class of solutions to the Einstein-Euler
equations that have a first post-Newtonian expansion. Using the
methods here, this can be improved to the
second post-Newtonian order. We will report
on this in a separate article.

In this article, we follow the approach of \cite{Oli06} to analyze
the limit $\ep\searrow 0$ of solutions to the Einstein-Euler equations. This requires that we
replace the metric $g_{ij}$ and fluid velocity
$v^i$ with new variables that are compatible with the limit
$\ep \searrow 0$. The new gravitational variable is a density $\ufb^{ij}$
defined via the formula
\leqn{metrecA}{
g^{ij} = \frac{\ep}{\sqrt{-\det(Q)}}Q^{ij}
}
where
\leqn{metrecB}{
Q^{ij} = \begin{pmatrix} \delta^{IJ} & 0 \\ 0 & 0 \end{pmatrix}
+  \ep^2 \begin{pmatrix} 4 \ufb^{IJ} & 0 \\ 0 & -1 \end{pmatrix}
+ 4\ep^3 \begin{pmatrix} 0 & \ufb^{I4} \\ \ufb^{J4} & 0 \end{pmatrix}
+ 4 \ep^4\begin{pmatrix} 0 & 0 \\ 0 & \ufb^{44} \end{pmatrix} .
}
From these formulas, it not difficult to see that the density $\ufb^{ij}$ is equivalent to the metric $g_{ij}$
for $\ep > 0$, and is well defined at $\ep =0$.
For the fluid, a new velocity variable $w^i$ is defined by
\leqn{wdef.intro}{
v^I = w^I \AND w^4=\frac{v^4-1}{\ep}\, .
}

For technical reasons, we assume an isentropic equation of
state
 \leqn{eos}{ p =
K\rho^{(n+1)/n} } for the fluid
where $K \in \Rbb_{>0}$, $n\in \Nbb$. This
allows us to use
 a technique of Makino \cite{Mak} to regularize the fluid
equations by the use of the fluid density variable $\alpha$ defined by
\leqn{dendef}{ \rho = \frac{1}{\bigl(4Kn(n+1)\bigr)^n}\alpha^{2n}. }
The resulting system can be put into
a symmetric hyperbolic system that
is regular across the
fluid-vacuum interface.
In this way,
it is possible to construct solutions to the Einstein-Euler
equations that represent compact gravitating fluid bodies (i.e. stars)
both in the Newtonian and
relativistic setting \cite{Mak,Ren92}.

The main point of introducing the gravitational-matter variables
$\{\ufb^{ij},w^i,\alpha\}$ is that in a harmonic gauge
the Einstein-Euler equations can be cast into
a singular  (non-local) symmetric hyperbolic system of the form
\leqn{EFsym2.intro}{ b^0(\epsilon W)\partial_t W =
\frac{1}{\epsilon}c^I\partial_I W + b^I(\epsilon,W)\partial_I W +
F(\epsilon,W).}
As shown in \cite{Oli06}, for appropriately chosen initial data this form is suitable
to derive $\ep$-independent energy estimates
that can be use to analyze the behavior
of the solutions as $\ep\searrow 0$, and extract a Newtonian limit. We note
that singular hyperbolic systems of the form \eqref{EFsym2.intro} have
been  extensively studied
\cite{BK,KM82,Kreiss,Scho86,Scho88}, but, as discussed in \cite{Oli06}, are
not directly applicable to the Einstein-Euler equations due to
initial data that does not lie in the
standard Sobolev space $H^k(\Rbb^3)$.

For general initial data, the $\ep$-independent
energy estimates from \cite{Oli06} are not enough to control
the solution in the limit $\ep\searrow 0$. In this paper, we show that
when the energy estimates are used in conjunction with dispersive estimates for the wave equation  a
larger class of initial data can be chosen so that the resulting solutions
still have a Newtonian limit. However, unlike the situation in \cite{Oli06}, the gravitational
variables do not converge to an $\ep$-independent limit. Instead, they
converge to a solution of a singular $\ep$-dependent wave equation. Following the terminology
used in other singular hyperbolic problems \cite{Scho88}, we refer to this type of limit as a
fast limit.
We note that
dispersive wave estimates have been used previously in a similar fashion to analyze
the (singular) incompressible limit for the Euler equations \cite{Iso87,Ukai86}.

The precise statement of the existence of a Newtonian limit
is contained in the following Theorem which is the main result of this article.
A proof can be found in section \ref{nlim}. A definition of the weighted spaces
$H^k_{\delta,\ep}$ $(H^k_{\delta}:=H^k_{\delta,1})$ can be found in Appendix A
of \cite{Oli06}. We also define $X_{T,s,k,\delta} := \cap^{s+1}_{\ell=0} C^\ell([0,T),H^{k-\ell}_\delta)$.
\begin{thm} \label{mthm} \mnote{[mthm]}
Suppose $-1 < \delta < -1/2$, $s\in \mathbb{Z}_{\geq 2}$, $R>0$,
$k\in \mathbb{Z}_{\geq 3+s}$, $\alphao,\wo^I \in H^k_{\delta-1}$,
$\supp\, \alphao \subset B_R$, $\mathfrak{z}^{IJ}\in H^{k+1}_\delta$,
$\mathfrak{z}^{IJ}_4 \in H^k_{\delta-1}$. Then there exists
a $T>0$, $\ep_0>0$, and maps
\alin{mth1}{
&\ufb_\ep^{ij}(t) \quad : \quad \ufb_\ep^{ij}(t)-\ufb^{ij}(0),\; \del_I \ufb^{ij}_\ep(t),\; \del_t\ufb^{ij}_\ep(t) \in X_{T,s,k,\delta-1}\quad 0<\ep \leq \ep_0, \\
&\ufbt_\ep^{ij}(t) \quad : \quad \ufbt_\ep^{ij}(t)-\ufb^{ij}(0),\; \del_I \ufbt^{ij}_\ep(t),\; \del_t\ufbt^{ij}_\ep(t) \in X_{T,s,k,\delta-1}
\quad 0< \ep \leq  \ep_0,\\
&\rho_\ep(t),\; w^i_\ep(t) \in X_{T,s,k,\delta-1}  \quad  0 < \ep \leq \ep_0 ,\\
& \rhot(t),\; \wt^I(t) \in X_{T.s,k,\delta-1}, \quad  \Phit(t)\in X_{T,s,k+2,\delta}\quad
\text{with} \quad \del_t\Phit(t)\in X_{T,s,k+1,\delta-1} ,
}
such that
\begin{itemize}
\item[(i)] the triple $\{\ufb^{ij}_\ep(x,t),\rho_\ep(x,t),w^i_\ep(x,t)\}$ determines
a solution
to the Einstein-Euler equations \eqref{EEeqn} in the harmonic gauge for $0<\ep\leq \ep_0$ on
the spacetime region $(x^I,t=x^4) \in D=\Rbb^3\times [0,T)$ with ADM mass given by
\eqn{mth2}{
m_{\text{\rm ADM}} = \int_{\Rbb^3}\rhot(x,0)\, dx^3 + \text{\rm O}(\ep^2),
}
\item[(ii)] $\{\rhot(x,t),\wt^I(x,t),\Phit(x,t)\}$ is a solution
to the Poisson-Euler equations \eqref{newtPE.1}-\eqref{newtPE.3} with
initial data $\rhot|_{t=0} = (4Kn(n+1))^{-1}\alphao^{2n}$, $\wt^I|_{t=0} =\wo^I$,
\item[(iii)] $\ufbt_\ep(x,t)$ is a solution to the wave equation
\eqn{mthb}{
\ep^2\del_t^2 \ufbt_\ep^{ij} - \Delta \ufbt_\ep^{ij} = -\delta^i_4\delta^j_4 \rhot +\ep^2\delta^i_4\delta^j_4\del_t^2\Phit,
}
with initial conditions
\alin{mth3b}{
\ufbt^{ij}_\ep\bigl|_{t=0} & = \delta^i_I\delta^j_J\zf^{IJ}
-2\Delta^{-1}\del_I\zf_4^{IJ}\delta^{(i}_4\delta^{j)}_J + \delta^{i}_4\delta^j_4
(\Phit\bigl|_{t=0}+\Delta^{-1}\del^2_{IJ}\zf^{IJ}), \\
\del_t\ufbt^{ij}_\ep\bigl|_{t=0} & = \frac{1}{\ep} \bigl(\delta^i_I\delta^j_J\zf_4^{IJ}
-2\del_I\zf^{IJ}\delta^{(i}_4\delta^{j)}_J + \delta^{i}_4\delta^j_4
\Delta^{-1} \del^2_{IJ}\zf_4^{IJ} \bigr) + \delta^i_4\delta^j_4\del_t\Phit\bigl|_{t=0} ,
}
\item[(iv)]
\eqn{mth4}{\norm{\rho_\ep(t)-\rhot(t)}_{H^{k-2}} +
\norm{w^I_\ep(t)-\wt^I(t)}_{H^{k-2}} + \norm{w^4_\ep(t)}_{H^{k-2}} \lesssim \ep,
}
and
\eqn{mth5}{\norm{\ufb^{ij}_\ep(t)-\ufbt^{ij}_{\ep}}_{L^6}+
\norm{\del_I\ufb^{ij}_\ep(t)- \del_I\ufbt^{ij}_{\ep}(t)}_{H^{k-2}} +
\norm{\ep\del_t \ufb^{ij}_\ep(t)- \ep\del_t\ufbt^{ij}_{\ep}(t)}_{H^{k-2}}
\lesssim \ep }
for all $(t,\ep) \in [0,T)\times (0,\ep_0]$.
\end{itemize}
\end{thm}
From the above theorem, the interpretation of the limiting solution
is clear. The $\{\rhot,\wt^I\}$ satisfies the standard Poisson-Euler equations of Newtonian
gravity with the obvious interpretation as the fluid density and three-velocity, while the $\ufbt^{ij}_\ep$ represent high frequency gravitational radiation propagating on
a flat background with the fluid density and Newtonian potential acting as source terms.

%% file: red.tex
\sect{red}{Reduced Einstein Equations}

To aid in deriving the appropriate symmetric hyperbolic system for
the gravitational variables, we temporarily introduce a new set of
coordinates related to old ones by the simple rescaling
\eqn{bcoords}{ \xb^J = x^J , \quad \xb^4 = x^4/\epsilon, } and let
\eqn{bpartial}{
\partial_i = \frac{\partial\;}{\partial x^i} \, ,
\quad \delb_i =  \frac{\partial\;}{\partial \xb^i} \, . } In
the new coordinates, the metric $\gb_{ij}$ and its inverse
$\gb^{ij}$ are given by \leqn{bmetricdef}{ (\gb_{ij}) =
\begin{pmatrix}
g_{IJ} & \ep g_{I4} \\
\ep g_{4J} & \ep^2 g_{44}
\end{pmatrix}
\quad \text{and} \quad
(\gb^{ij}) = \begin{pmatrix}
g^{IJ} & \ep^{-1} g^{I4} \\
\ep^{-1} g^{4J} & \ep^{-2} g^{44}
\end{pmatrix}\, .
} Next, we consider the metric density \leqn{densdef}{ \gfb^{ij} =
\sqrt{|\gb|}\, \gb^{ij} \quad \text{where} \quad |\gb| =
-\det(\gb_{ij})\, . }
We note that the metric $\gb^{ij}$ is related to the density
$\gfb^{ij}$ by the following formula
\leqn{den2metb}{ \gb^{ij} = \frac{1}{\sqrt{|\gb|}}\gfb^{ij} \quad
\text{where} \quad |\gb| = -\det{\gfb^{ij}}\, , } and hence
\leqn{den2met}{ (g^{ij}) = \frac{1}{\sqrt{|\gb|}}
\begin{pmatrix}
\gfb^{IJ} & \ep \gfb^{I4} \\
\ep \gfb^{4J} & \ep^2 \gfb^{44}
\end{pmatrix} \,.
}
To obtain a gravitational variable that is regular and non-trivial
in the limit $\ep \searrow 0$, we define \leqn{udensdef}{
\ufb^{ij} := \frac{1}{4\epsilon^2}\bigl(\gfb^{ij}-\eta^{ij}\bigr) }
where \eqn{minkowski}{ (\eta^{ij}) = \begin{pmatrix}
\id_{\!\!3\times 3} & 0 \\
0 & -1
\end{pmatrix}
} is the Minkowski metric density.
As stated in the introduction, for $\ep > 0$,
the metric $g_{ij}$ can be recovered from the density $\ufb^{ij}$ via
the formulas \eqref{metrecA}-\eqref{metrecB}.
In the $(\xb^i)$ coordinate system, the Christofell symbols are
given by \leqn{ChristA}{ \Gammab^{k}_{ij} =
\ep^2\bigl(\gfb^{km}(2\gfb_{i\ell}\gfb_{jp} - \gfb_{ij} \gfb_{\ell
p})\delb_{m}\ufb^{\ell p} + 2( \gfb_{\ell
p}\delta^{k}_{(i}\delb_{j)}\ufb^{\ell p}-
2\gfb_{\ell(i}\delb_{j)}\ufb^{k\ell} )\bigr) \, . } These
are related to the  Christofell symbols in the $(x^i)$ coordinate system as follows
\lgath{ChristB}{
\Gamma^A_{44} = \ep^{-2} \Gammab^{A}_{44} \, , \quad
\Gamma^{4}_{44} = \ep^{-1} \Gammab^{4}_{44} \, ,
\quad \Gamma^{4}_{A4} = \Gammab^{4}_{A4}\, ,\label{ChristB.1}\\
\quad \Gamma^{4}_{AB} = \ep \Gamma^{4}_{AB}\, ,\quad \Gamma^A_{B4}
= \ep^{-1}\Gammab^A_{B4} \AND \Gamma^{A}_{BC} = \Gammab^{A}_{BC}
\, .\label{ChristB.2} }

Using \eqref{ChristA}, a straightforward calculation shows that the
Einstein tensor $\Gb^{ij}$ is given in terms of the density
$\ufb^{ij}$ by \leqn{Gb}{ \Gc^{ij}:= \frac{1}{2\ep^2}|\gb|\,
\Gb^{ij} = \gfb^{k\ell}\delb^2_{k\ell} \ufb^{ij}+
\ep^2\bigl(A^{ij} + B^{ij} + C^{ij}\bigr) + D^{ij} } where
\lalign{Gbdef}{
|\gb| & = -\det(\gfb^{ij})\, , \label{Gbdef.1} \\
A^{ij} & = 2\bigl(\Half \gfb_{k\ell} \gfb_{mn} - \gfb_{km}
\gfb_{\ell n} \bigr) \bigl(\gfb^{ip} \gfb^{jq} - \Half \gfb^{ij}
\gfb^{pq} \bigr)\delb_{p} \ufb^{k\ell}\delb_{q}\ufb^{mn}
\label{Gbdef.2}\, ,\\
B^{ij} & = 4\gfb_{k\ell}\bigl(2\gfb^{n(i }
\delb_{m}\ufb^{j)\ell}\delb_{n}\ufb^{k m} - \Half
\gfb^{ij}\delb_{m}\ufb^{k n}\delb_{n}\ufb^{m \ell}
-\gfb^{mn}\delb_{m}\ufb^{ik}\delb_{n}\ufb^{j\ell}\bigr)\label{Gbdef.3} \, , \\
C^{ij} & = 4\bigl(\delb_k\ufb^{ij}\delb_{\ell}\ufb^{k\ell}-
\delb_k\ufb^{i\ell}\delb_\ell\ufb^{jk})\label{Gbdef.4} \, , \\
D^{ij} & := \gfb^{ij}\delb^2_{k\ell} \ufb^{k\ell} - 2
\delb^2_{k\ell}\ufb^{k(i}\gfb^{j)\ell}\, . \label{Gbdef.5} }
To
fix the gauge,  we assume that \leqn{harm}{
\delb_{i}\ufb^{ij}
= 0.} For $\ep > 0$, this is equivalent to
to the harmonic gauge \leqn{harmB}{ \del_{i}\gf^{ij} =
\del_{i}\bigl(\sqrt{-\det(g_{k\ell})}\, g^{ij} \bigr) = 0. }

Setting \leqn{Gred}{ \Gcr^{ij} :=\Gc^{ij}-D^{ij} =
\gfb^{k\ell}\delb^2_{k\ell}\ufb^{ij} +
\epsilon^2\bigl(A^{ij}+B^{ij}+C^{ij}\bigr)  } and \eqn{streseng}{
\Tc^{ij} := \ep^2 |\gb|\,\Tb^{ij} = |\gb|\begin{pmatrix}
\ep^2T^{IJ} & \ep^{1} T^{I4} \\
\ep^{1}T^{4J} & T^{44}
\end{pmatrix}
} the Einstein equations $G^{ij} = 2\ep^4 T^{ij}$ in the gauge
\eqref{harm} become \leqn{Einred}{ \Gcr^{ij}=  \Tc^{ij}. } We
will refer to these as the \emph{reduced Einstein equations}.

To write the reduced Einstein equations in first order form, we
introduce the variables \eqn{fo1}{ \ufb^{ij}_{k} :=
\delb_{k}\ufb^{ij} = \left\{ \begin{array}{ll}
\partial_{I}\ufb^{ij} & \text{if $k=I$}\\
\ep \partial_{4} \ufb^{ij} & \text{if $k=4$}
\end{array} \right. \, .
} The reduced Einstein equations then become \alin{fo2}{
-\gfb^{44}\delb_4 \ufb^{ij}_4 & =
2\gfb^{4I}\delb_I\ufb^{ij}_4 + \gfb^{IJ}\delb_I\ufb^{ij}_J
 +\ep^2\bigl(A^{ij}+B^{ij}+C^{ij}) - \Tc^{ij} \, ,\\
\gfb^{IJ}\delb_4\ufb^{ij}_J & = \gfb^{IJ}\delb_{J}\ufb^{ij}_4\, , \\
\delb_4 \ufb^{ij} &= \ufb^{ij}_4 \, , } or equivalently
\alin{fo3}{ -\gfb^{44}\partial_4 \ufb^{ij}_4 & =
\frac{2}{\epsilon}\gfb^{4I}\partial_I\ufb^{ij}_4 +
\frac{1}{\epsilon}\gfb^{IJ}\partial_I\ufb^{ij}_J
+\ep\bigl(A^{ij}+B^{ij}+C^{ij}) -\frac{1}{\epsilon} \Tc^{ij} \, ,\\
\gfb^{IJ}\partial_4\ufb^{ij}_J & =
\frac{1}{\ep}\gfb^{IJ}\partial_{J}\ufb^{ij}_4\, , \\ \partial_4
\ufb^{ij} &= \frac{1}{\ep}\ufb^{ij}_4 \, . } Next, we define
\leqn{fo4}{ \uf^{ij} := \ep \ufb^{ij}, \quad
\uf^{ij}_k := \ufb^{ij}_k ,  } and let
\eqn{Vcaldef}{ \Vc = \{ \, (r^{ij}) \in \Mbb_{4\times
4} |\; \det (\eta^{ij} + 4 r^{ij}) > 0 \, \}\, . } Then using
vector notation \eqn{fo5}{ \ufv^{ij} := ( \uf_4^{ij}, \uf_J^{ij},
\uf^{ij})^T \, , } the reduced Einstein equations take the form
\leqn{fo6}{ A^4 (\ep\ufv)\partial_4\ufv^{ij} =
\frac{1}{\ep}C^{I}\partial_I\ufv^{ij} +
A^I(\ufv)\partial_I\ufv^{ij} + \Fb^{ij}(\ep,\ufv)
-\frac{1}{\ep}(\Tc^{ij},0,0)^T, } where
\leqn{fo7}{ A^{4}(\ep\ufv) =
\begin{pmatrix}
1-4\ep\uf^{44} & 0 & 0\\
0 & \delta^{IJ}+4\ep\uf^{IJ} & 0 \\
0 & 0 & 1
\end{pmatrix},
}
\leqn{fo8}{ C^{I} = \begin{pmatrix}
0 & \delta^{IJ} & 0 \\
\delta^{IJ} & 0 & 0 \\
0 & 0 & 0
\end{pmatrix},
} \leqn{fo9}{ A^{I}(\ufv) = \begin{pmatrix}
8\uf^{4I} & 4\uf^{IJ} & 0 \\
4\uf^{IJ} & 0 & 0 \\
0 & 0 & 0
\end{pmatrix},
}
\leqn{fo10aa}{
\Fb_0^{ij}(\ufv) = (0,0,\uf^{ij}_4)^T,
}
and \leqn{fo10}{ \Fb^{ij}(\ufb,\ep\ufv) = (A^{ij}+B^{ij}+C^{ij},0,0)^T
.
}
Here we are using the notation \eqn{fo10a}{ \uf =
(\uf^{ij}) \AND \uf_k = (\uf^{ij}_k) \, . }

The stress-energy
tensor is given in terms of the $\uf$ variable by \lalign{fo11a}{
(T^{ij}) = \rho (v^i v^j) &+\frac{1}{\sqrt{|\gb|}}\begin{pmatrix}
\delta^{IJ} p & 0 \\
0 & 0
\end{pmatrix}
+ \frac{\ep}{\sqrt{|\gb|}}\begin{pmatrix}
4\uf^{IJ} p & 0 \\
0 & 0
\end{pmatrix} \notag \\
& + \ep^2\left( p(v^i v^j) + \frac{p}{\sqrt{|\gb|}}\begin{pmatrix}
0 & 4\uf^{I4} \\
4\uf^{4J} & -1+4\ep\uf^{44}
\end{pmatrix}
 \right), \label{fo11}
} which we can write as \leqn{fo12}{ \frac{1}{\ep}(\Tc^{ij}) =
\begin{pmatrix}
0 & 0 \\
0 & \ep^{-1}\rho
\end{pmatrix}
+ \Sc^{ij} } where \lalign{fo13}{ (&\Sc^{ij})  = \rho
\begin{pmatrix}
0 & |\gb|v^{I}v^{4} \\
|\gb|v^{J}v^{4}  & \ep^{-1}\bigl[(|\gb|-1)(v^{4})^2 +((v^4)^2-1)\bigr]
\end{pmatrix} +\notag\\
& \ep |\gb|\begin{pmatrix} (\rho+\ep^2 p) v^{I}v^{J} +
|\gb|^{-1/2}p(\delta^{IJ}+4\ep\uf^{IJ}) &
\ep pv^{I}v^{4} +  4 \ep |\gb|^{-1/2}p\uf^{I4} \\
\ep pv^{J}v^{4} +  4 \ep |\gb|^{-1/2}p\uf^{4J} & p (v^4)^2 +
|\gb|^{-1/2}p(-1+4\ep\uf^{44})
\end{pmatrix}.\label{fo13.1}
}

Letting (see \eqref{wdef.intro}) \leqn{fluvars}{ \wv = (\alpha,w^i)^T, }
we
can decompose $\Sc^{ij}$ as
\leqn{fo14}{ \Sc^{ij} = \Sc_0^{ij} + \ep \Sc_1^{ij}, } where
\lalign{fo15.1}{ \Sc_0^{ij}&(\uf,\wv,\ep\uf,\ep \wv)  = \notag \\
&\rho
\begin{pmatrix} 0 & |\gb|w^I(1+\ep w^4) \\ |\gb| w^J(1+\ep w^4) &
\ep^{-1}\bigl[  (|\gb|-1)(1+\ep w^4)^2 + \bigl((1+\ep w^4)^2 -1
\bigr) \bigr]
\end{pmatrix}, \label{fo15}
} and \lalign{fo16.1}{ \Sc_1^{ij}&(\wv,\ep\uf,\ep\wv) = \notag \\
&|\gb|\begin{pmatrix} \rho w^I w^J+p\ep w^I \ep
w^J+|\gb|^{-1/2}p\gfb^{IJ} & p \ep w^I(1+\ep
w^4)+4|\gb|^{-1/2}p\ep\uf^{I4} \\
p \ep w^J(1+\ep w^4)+4|\gb|^{-1/2}p\ep\uf^{J4} & p(1+\ep
w^4)^2+|\gb|^{-1/2} p (-1+4\ep \uf^{44})
\end{pmatrix}.\label{fo16} }

%% file: eul.tex
\sect{eul}{Regularized Euler equations}

In the coordinates $(\xb^i)$, the Euler equations are given by
\leqn{eul1}{ \nablab_i \Tb^{ij} =0} where $\Tb^{ij} = (\rho + \ep^2
p) \vb^i\vb^j + p \gb^{ij}$ and the fluid velocity $\vb^i$ is
normalized according to
\leqn{eul3}{\vb_i\vb^i = -\frac{1}{\ep^2}\, .}
To write \eqref{eul1} as a symmetric hyperbolic
system, we follow \cite{BrKa07} and differentiate \eqref{eul3} to get
\leqn{eul4}{\vb_i \nablab_j \vb^i = 0 \AND \vb^{j}\vb_i \nablab_j \vb^i = 0\, .}
Writing out
\eqref{eul1} explicitly, we have
\leqn{eul6}{(\delb_i\rho
+\ep^2\delb_i p)\vb^i\vb^j + (\rho+\ep^2 p)(\vb^j\nablab_i\vb^i
+\vb^i\nablab_i\vb^j) + \gb^{ij}\delb_i p = 0\, . } The operator
\eqn{proj}{L^j_i = \delta^j_i + \ep^2 \vb^j\vb_i} projects into
the subspace orthogonal to the fluid velocity $\vb^i$, i.e.
$L^j_{i}L^i_k = L^j_k $ and $L^j_i\vb^i=0$.  Using $L^j_k$ to
project the Euler equations \eqref{eul6} into components parallel
and orthogonal to $\vb^i$ yields, after using the relations
\eqref{eul3}-\eqref{eul4}, the following system \lgath{eul9}{
\vb^i\delb_i
\rho +(\rho+\ep^2 p)L^i_j\nablab_i\vb^j = 0 \, ,\label{eul9.1} \\
M_{ij}\vb^k\nablab_k \vb^j + \frac{1}{\rho+\ep^2 p} L^i_j \delb_i
p = 0\, ,\label{eul9.2}} where \eqn{Mdef}{ M_{ij} = \gb_{ij} +
2\ep^2 \vb_i\vb_j\, . }

As discussed in the introduction,
we use a Makino density variable
$\alpha$ (see \eqref{dendef} ) to regularize the
fluid equations in regions where the density and pressure vanish.
After multiplying
\eqref{eul9.1} by the square of the function
\eqn{hdef}{
h(\ep\alpha) = \left( 1 + \frac{1}{4n(n+1)}(\ep\alpha)^2\right)\,
,} a short calculation shows that the Makino density $\alpha$ and the fluid
four-velocity $\vb^i$ satisfy
\lgath{eul10}{ h^2\vb^i\delb_i
\alpha + q L^i_j\nablab_i\vb^j = 0 \, ,\label{eul10.1} \\
M_{ij}\vb^k\nablab_k \vb^j + q L^j_i \delb_j \alpha = 0\, , \label{eul10.2}
} where
\eqn{sound}{ s^2 = \frac{dp}{d\rho} =
\frac{1}{4n^2}\alpha^2} is the square of the speed of sound, and
\eqn{eul12}{ q = \frac{1}{2n
h(\ep\alpha)}\alpha\, .}
Instead of solving \eqref{eul10.1}-\eqref{eul10.2}, we consider the
following modified system
\lgath{eul10a}{ h^2\vb^i\delb_i
\alpha + q L^i_j\nablab_i\vb^j = 0,\label{eul10a.1} \\
M_{ij}\vb^k\nablab_k \vb^j + q L^j_i \delb_j \alpha
+ (\chi_{4\Rb}-1)M_{ij}\Gammab^j_{k\ell}\vb^k\vb^\ell = 0, \label{eul10a.2}
}
Here we are using
\eqn{chidef1}{\chi_\lambda(x) := \chi(x/\lambda) \qquad
\lambda>0,
}
where $\chi \in C^\infty(\Rbb^3)$ is a smooth cutoff function satisfying
$\chi(x) = 1$ for $|x|\leq 1$, $\chi(x) =0$ for $|x|\geq 2$,
and $0\leq \chi(x)\leq 1$ for all $x\in \Rbb^3$.

Since $w^I =\vb^I$ and $w^4 = \vb^4-1/\ep$, we can write \eqref{eul10a.1} and
\eqref{eul10a.2} as \leqn{eul13}{a^4 \del_4 \wv = a^I\del_I\wv + b}
where \lalign{adef}{ a^4 =&
\begin{pmatrix}
h^2(1+\ep w^4) & \ep q L^4_j \\
\ep q L^4_i & M_{ij}(1+\ep w^4)
\end{pmatrix} ,\label{adef.1} \\
a^I = & \begin{pmatrix}-h^2w^I & -q L^I_j \\
-q L^I_i & -M_{ij}w^I\end{pmatrix} , \label{adef.2}
\intertext{and}
b =& \begin{pmatrix} -q L^i_j \Gammab^j_{i\ell}\vb^\ell\\
-\chi_{4\Rb}M_{ij}\Gammab^j_{k\ell}\vb^k\vb^\ell \end{pmatrix} .
\label{adef.3}} From \eqref{den2metb}, \eqref{udensdef},
\eqref{fo4}, and \eqref{fo14}, we find that \leqn{metexp}{
\gb_{ij} = \eta_{ij} + f_{ij}(\ep\uf) } where the $f_{ij}(y)$ are
analytic and satisfy $f_{ij}(y) = \text{O}(|y|)$ as $y\rightarrow
0$, while  \eqref{ChristA} shows that
\leqn{Christexp}{\Gammab^k_{ij}
=\ep\bigl[\eta^{km}\bigl(2\eta_{i\ell}\eta_{jp}-\eta_{ij}\eta_{\ell
p}\bigr)\ep\uf^{lp}_m + 2\bigl(\eta_{\ell p}\delta^k_{(i}\ep\uf^{\ell
p}_{j)}-2\eta_{\ell(i} \ep \uf^{k\ell}_{j)}\bigr)\bigr] + \ep
f^{k}_{ij}(\ep\uf,\ep\uf_m) } for functions $f^k_{ij}(\ep \uf, \ep
\uf_m)$ that are analytic for $\ep\uf \in \Vc$, linear in the $\ep
\uf_m$, and satisfy $f^k_{ij}(0,y)=0$. The expansion \eqref{metexp}
allows us to write
 \leqn{Mexp}{ M_{ij}
= \gb_{ij} + 2\ep^2\gb_{ik}\gb_{j\ell}\vb^k\vb^\ell= \delta_{ij} +
m_{ij}(\ep\uf,\ep w^k),} and \leqn{Lexp}{L^j_i = \delta^j_i + \ep^2
\gb_{ik}\vb^k\vb^j = \delta^j_i -\delta_i^4\delta^{j}_{4} +
\ell^j_i(\ep\uf,\ep w^k) } for functions $\ell^j_i(\ep\uf,\ep w^k)$ and
$m_{ij}(\ep\uf,\ep w^k)$ that satisfy
$\ell^j_i(0,0)=m_{ij}(0,0) = 0$, and are analytic for $\ep\uf \in \Vc$. Using
\eqref{metexp}-\eqref{Lexp}, we can express the $a^i$ and $b$
as \lalign{aexp}{ a^4 &= \begin{pmatrix} 1 & 0 \\ 0
& \delta_{ij}\end{pmatrix} + \ah^4(\ep\uf,\ep \wv) ,
\label{aexp.1} \\
a^I &= \begin{pmatrix} -w^I & -\frac{\alpha}{2n}\delta^I_j \\
-\frac{\alpha}{2n}\delta^I_i & -\delta_{ij}w^I
\end{pmatrix} +  w^I\ah(\ep\uf,\ep\wv) +
\alpha\ah^I(\ep\uf,\ep\wv) , \label{aexp.2}
\intertext{and}
b &= \begin{pmatrix} 0 \\
\chi_{4\Rb}\left[-\eta^{im}\bigl(2\eta_{4\ell}\eta_{4p}+\eta_{\ell
p}\bigr)\uf^{lp}_m - 2\bigl(\eta_{\ell p}\delta^i_4\uf^{\ell
p}_{4}-2\eta_{\ell 4}\uf^{i\ell}_{4}\bigr) \right]
\end{pmatrix} \notag  \\
& \qquad +\begin{pmatrix}\alpha \bh_1(\ep\uf,\ep\wv)\cdot
\ep\uf_k\\
\chi_{4\Rb}\bh_2(\ep\uf,\ep\wv)\cdot \uf_k\end{pmatrix}. \label{aexp.3}} We observe
that the matrices  $\ah^4$,
$\ah$, and $\ah^I$ are symmetric,
and the maps $\ah^4$, $\ah$, $\ah^I$, $\bh_1$, and $\bh_2$ are analytic
(for $\ep\uf \in \Vc$) and satisfy  $\ah^4(0,0)=0$,
$\ah^I(0,0) = 0$, $\ah(0,0)= 0$, $\bh_1(0,0)= 0$, and
$\bh_2(0,0)=0$. This shows that the system \eqref{eul13} is
\emph{symmetric hyperbolic} on a region where $(\ep\uf,\ep \wv)$
is small enough to ensure that $a^4$ is positive definite. This
can always be arranged by taking $\ep$ small enough and since we
are interested in the limit $\ep \searrow 0$ no generality is lost
by assuming this.

%% file: uniform.tex
\sect{uniform}{Uniform local existence}

The combined systems \eqref{fo11} and \eqref{eul13} can be written
as \lalign{EFsym1.1}{ b^0(\ep V,\ep^2 U)\del_t V = &
\frac{1}{\ep}c^I\del_I V + b^I(V,\ep U,\ep V,\ep^2 U)\del_I V + \notag \\
& f_0(V,\ep U,\ep V,\ep^2 U) + \ep f_1(V,\ep U,\ep V,\ep^2 U)+
\frac{1}{\ep}g(V), \label{EFsym1}} where
\lalign{EFsym2}{ U & = ( 0 , 0 ,\ufbo^{ij}, 0, 0 )^T,
\qquad\qquad \ufbo^{ij} = \ufb^{ij}\bigl|_{t=0}, \label{EFsym2.1} \\
V & = ( \uf^{ij}_{4} , \uf^{ij}_{J} , \delta\uf^{ij} , \alpha ,
w^i )^T\, ,  \qquad \qquad \delta\uf^{ij}=
\uf^{ij}-\ep \ufbo^{ij} , \label{EFsym2.2} \\
b^0(\ep V,\ep^2 U) & = \begin{pmatrix} A^4(\ep\uf) & 0 \\
0 & a^4(\ep \uf,\ep \wv)  \end{pmatrix} , \label{EFsym2.3} \\
c^I & = \begin{pmatrix}
C^I & 0 \\
0 & 0
\end{pmatrix}
\label{EYsym2.4},\\
b^I(V,\ep U,\ep V,\ep^2 U) & = \begin{pmatrix}  A^I(\uf) & 0 \\
0 & a^I(\wv,\ep \uf, \ep \wv) \end{pmatrix}  , \label{EFsym2.5}  \\
f_0(V,\ep U,\ep V,\ep^2 U) & = \begin{pmatrix}
\Fb_0^{ij}(\ufv)-\Sc_0^{ij}(\uf,\wv,\ep\uf,\ep \wv) \\
b(\ufv,\wv,\ep\ufv,\ep\wv)
\end{pmatrix}    , \label{EFsym2.6} \\
f_1(V,\ep U,\ep V,\ep^2 U) & = \begin{pmatrix}
\Fb_1^{ij}(\ufv,\ep\ufv)-\Sc_1^{ij}(\wv,\ep\uf,\ep \wv) \\
0
\end{pmatrix} \, ,\label{EFsym2.7} \intertext{and}
g(V) & = ( -\delta^{i}_4\delta^{j}_4 \rho(\alpha),
0,\ldots,0)^T  . \label{EFsym2.8}
}
For initial data, we will often use the notation \eqn{t0}{
\underset{o}{z} = z|_{t=0}\, .}

In addition to solving these evolution equations, we must also solve the
following constraint equations on the initial
hypersurface $\Sigma$ $=$ $\{(x^I,0)\,|\, (x^I)\in \Rbb^3 \}$
to get a full solution to the Einstein-Euler equations:
\lalign{con}{
\Cc^j & := \Gc^{4i}- \Tc^{4i} = 0 \qquad \text{(gravitational constraint equations),} \label{con.1}\\
\Hc^j & := \delb_i\ufb^{ij} = 0 \qquad \text{(harmonic gauge condition),} \label{con.2}\intertext{and}
\Nc & := \ep\vb_i\vb^i + \frac{1}{\ep} = 0 \qquad  \text{(fluid velocity normalization).}\label{con.3}
}

To fix a  region on which the system where both the evolution \eqref{EFsym1}
and constraint equations \eqref{con.1}-\eqref{con.3} are well
defined, we note from \eqref{fo6}, \eqref{aexp.1}, and the
invertibility of the Lorentz metric $(\eta^{ij})$ that there
exists a constant $K_0>0$ such that \leqn{contA1}{ -\det(\eta^{ij}
+ 4\ep\uf^{ij}) > 1/16\, , \quad 1+\ep w^4
> 1/16\, ,} \leqn{contA2}{ \quad A^4(\ep \uf) \geq \frac{1}{16} \id\, ,
\quad a^4(\ep\uf,\ep \wv) \geq \frac{1}{16}\id, } and
\leqn{contA3}{ |A^4(\ep \uf)|\leq 16 \, , \quad |a^4(\ep\uf,\ep
\wv)| \leq 16} for all $|\ep\uf|\leq 2K_0$, $|\ep
w^i|\leq 2K_0$, $|\ep \alpha| \leq 2K_0$. The choice of the bounds
$1/16$ and $16$ is somewhat arbitrary, and they can be replaced by
any number of the form $1/M$ and $M$ for any $M > 1$ without
changing any of the arguments presented in the following sections.
However, since we are interested in the limit $\ep \searrow 0$, we
lose nothing by assuming $M=16$.

\subsect{idat}{Newtonian initial data}

To generate a one parameter family of solutions to the constraint equations \eqref{con.1}-\eqref{con.3}
that is regular in the limit $\ep\searrow 0$, we use a slight variation of
the method used in \cite{Oli06}, which is based on previous
work by Lottermoser \cite{Lott}.
Before we state the theorem, we note from \eqref{dendef}, \eqref{eos}, and
the weighted multiplication inequality (see \cite{Oli06} Lemma A.8 ) that if
$\alpha\in H^k_{\delta}$ $(\delta \leq 0,k>3/2)$ then
$\rho,p \in H^{k}_{\delta}$.
\begin{prop}\label{idatA} \mnote{[idatA]}
Suppose $-1<\delta < 0$, $k > 3/2+1$,  $R>0$
and
$(\tilde{\rho},\tilde{p},\tilde{w}^I,\tilde{\zf}_4^{IJ},\tilde{\zf}^{IJ})
 \in (H^{k-2}_{\delta-2})^2\times H^{k}_{\delta-1}\times
H^{k-1}_{\delta-1}\times B_R(H^{k}_{\delta}) \,.$
Then there exists an
$\ep_0>0$, an open neighborhood $U$ of
$(\tilde{\rho},\tilde{p},\tilde{w}^I,\tilde{\zf}_4^{IJ},\tilde{\zf}^{IJ})$,
and analytic maps $(-\ep_0,\ep_0)\times U \rightarrow
H^{k}_{\delta-1}$ $\; : \;$
$(\ep,\rho,p,w^I,\zf_4^{IJ},\zf^{IJ})$$\mapsto$ $w^4$,
$(-\ep_0,\ep_0)\times U \rightarrow H^{k}_{\delta}$ $\; : \;$
$(\ep,\rho,p,w^I,\zf_4^{IJ},\zf^{IJ})$$\mapsto$ $\phi$,
$(-\ep_0,\ep_0)\times U \rightarrow H^{k}_\delta$ $\; : \;$
$(\ep,\rho,p,w^I,\zf^{IJ}_4,\zf^{IJ})$ $\mapsto$ $\wf^I$ such
that for each $(\rho,p,w^I,\zf^{IJ}_4,\zf^{IJ})\in U$,
$(\ep,\rho,p,w^I,w^4,\ufb^{ij}_4,\delb_4\ufb^{ij})$ is a solution to
the three constraints \eqn{idatA2}{ \Cc^j=0\, ,\quad  \Hc^j=0\, ,
\AND \Nc =0\,,} where \alin{idatA1}{ (\ufb^{ij}) &
=\begin{pmatrix} \zf^{IJ} &
\wf^I \\
\wf^J &
\phi
\end{pmatrix}
,\\
(\ep \del_t\ufb^{ij}) & = \begin{pmatrix} \zf^{IJ}_4 & -
\del_{K}\zf^{KI}
\\ -\del_{K}\zf^{KJ}& -\del_K\wf^K
\end{pmatrix} \, ,
\intertext{and}
w^4 &=  -\frac{1}{\ep} +
\frac{-\ep\gb_{4J}w^J-\sqrt{\ep^2(\gb_{4J}w^J)^2-\gb_{44}
(\ep^2\gb_{IJ}w^Iw^J+1)}}{\ep \gb_{44}}\, .
}
Moreover, if we let
$\phi_0 = \phi|_{\ep=0}$, $\wf^I_0=\wf^I|_{\ep =0}$,
and $w^4_0 = w^4|_{\ep=0}$,
then $\phi_0$, $\wf^I_0$, and $w^4_0$ satisfy the equations \eqn{idatA3a}{
\Delta\phi_0 = \rho + \del^2_{IJ}\zf^{IJ}\, , \quad \Delta\wf^I_0 = -\del_L\zf^{LJ}_4
, \AND w^4_0 =  0 \, , } respectively.
\end{prop}
\begin{proof} The proof follows from a simple adaptation of the proof
of Proposition 5.1 in \cite{Oli06}.
\end{proof}

\begin{cor} \label{mass} \mnote{[mass]} For $-1<\delta<-1/2$, the
ADM mass of the 1-parameter family of initial data constructed
in Proposition \ref{idatA} satisfies
\eqn{mass1}{
m_{\text{\rm ADM}} =  \int_{\Rbb^3} \rho\, dx^3 +
\text{\rm O}(\ep) .
}
\end{cor}
\begin{proof}
For fixed
$(\rho,p,w^I,\zf_4^{IJ},\zf^{IJ})
 \in (H^{k-2}_{\delta-2})^2\times H^{k}_{\delta-1}\times
H^{k-1}_{\delta-1}\times H^{k}_{\delta}$, it follows
from Proposition \ref{idatA} that for $\ep_0$ small enough,
the maps
\leqn{mass3}{
[0,\ep_0) \ni \ep \longmapsto \gb_{ij} \in H^{k}_{\delta}
\AND [0,\ep_0) \ni \ep \longmapsto \ep\del_t\gb_{ij}\in H^{k-1}_{\delta-1}
}
are analytic (see \eqref{bmetricdef} and \eqref{udensdef}). Moreover,
a short calculation shows that
\lalign{mass4}{
\gb^\ep_{ij} & = \eta_{ij} + 2\ep^2\bigl(\eta_{kl}u^{k\ell}\eta_{ij}-2\eta_{ik}u^{k\ell}\eta_{\ell j}\bigr) + \text{\rm O}(\ep^3) \label{mass4.1},
\intertext{and}
\ep\del_t\gb_{ij} &= 2\ep^2\bigl(\eta_{kl}u_4^{k\ell}\eta_{ij}-2\eta_{ik}u_4^{k\ell}\eta_{\ell j}\bigr) + \text{\rm O}(\ep^3), \label{mass4.2}
}
where
\lalign{mass6}{ (u^{ij}) &
=\begin{pmatrix} \zf^{IJ} &
\wf_0^I \\
\wf_0^J &
\phi_0
\end{pmatrix}, \label{mass6.1} \\
(u_4^{ij}) & = \begin{pmatrix} \zf^{IJ}_4 & -
\del_{K}\zf^{KI}
\\ -\del_{K}\zf^{KJ}& -\del_K\wf_0^K
\end{pmatrix}, \label{mass6.2}
\intertext{and}
\Delta\phi_0 &= \rho + \del^2_{IJ}\zf^{IJ}\, , \quad \Delta\wf^I_0 = -\del_L\zf^{LJ}_4.
\label{mass6.3}
}

Since $-1<\delta<-1/2$, it follows from Proposition 4.5 of \cite{Bart05} that
the total ADM energy-momentum $\Pbb=(\Pbb_j)$ for the initial data
$(\gb_{ij},\ep\del_t\gb_{ij})$ on the initial hypersurface $\Sigma = \{(x^I,0)\, |\,
(x^I)\in \Rbb^3\}$
can be calculated using the standard formulas
\lalign{mass7}{
\Pbb_4 &= -\frac{1}{4}\oint_{S_\infty}\bigl(\del^I\gb_{IJ}- \delta^{IK}\del_J \gb_{IK}\bigr)
\, dS^J, \label{mass7.1} \\
\Pbb_I &= \frac{1}{2}\oint_{S_\infty} \bigl( \bar{K}^{K}_K\delta_{IJ} - K_{IJ}\bigr)\, dS^J,
\label{mass7.2}
}
where the extrinsic curvature $\bar{K}_{IJ}$ is given by (see \eqref{ChristA})
\leqn{mass8}{
\bar{K}_{IJ} = -\sqrt{\frac{-1}{\gb^{44}}} \bar{\Gamma}^{4}_{IJ}.
}
Furthermore, the map
\leqn{mass9}{
[0,\ep_0) \ni \ep \longmapsto (\Pbb_j) \in \Rbb^4
}
is smooth by \eqref{mass3} and Theorem 5.1 of \cite{Bart05}.

By \eqref{ChristA}, \eqref{mass4.1}, \eqref{mass4.2}, and \eqref{mass8}, we find the following
expansions for the extrinsic curvature:
\leqn{mass10}{
\bar{K}_{IJ} = \ep^2\Bigl(2 u_4^{IJ} + \delta_{IJ}\bigl(u_4^{44}-\delta_{k\ell}u_4^{k\ell}\bigr)
+2\del_J u^{4I}+2\del_I u^{4J} \Bigr) + \text{\rm O}(\ep^3).
}
The smoothness of the map \eqref{mass9} and the two expansions \eqref{mass4.1} and \eqref{mass10} show
that the ADM energy-momentum can be expanded as
\lalign{mass11}{
\Pbb_4 &= -\ep^2\oint_{S_\infty}\bigl(\del^I\phi_0-\del_J\zf^{IJ}\bigr)\, dS_I
+\text{\rm O}(\ep^3), \label{mass11.1} \\
\Pbb_I &= \ep^2 \oint_{S_\infty} \bigl( \delta^{J}_I\del_K\wf_0^K - \del^J\wf_0^I
-\del_I\wf_0^{4J} -\zf_4^{IJ} \bigr)\, dS_J +\text{\rm O}(\ep^3).
\label{mass11.2}
}
Using the divergence theorem and \eqref{mass6.3}, the energy-momentum expansions \eqref{mass11.1} and \eqref{mass11.2}
simplify to
\leqn{mass12}{
\Pbb_4 = -\ep^2 \int_{\Rbb^3}\rho \, dx^3 +
\text{\rm O}(\ep^3) \AND
\Pbb_I = \text{\rm O}(\ep^3).
}
The proof of the corollary now follows from the above expansions
and the definition
\eqn{mass13}{
m_{\text{ADM}} = \frac{1}{\ep^2} \sqrt{\eta^{ij}\Pbb_i\Pbb_j}}
of the ADM mass.
\end{proof}

\subsect{eplim}{Uniform existence}

To prove local existence of solutions to
\eqref{EFsym1} on a uniform time interval independent of
$\ep$, we take the same approach as in \cite{Oli08} and
use a non-local modification of \eqref{EFsym1}.
The modified system is constructed as follows.
First, we replace $g(V)$ in
\eqref{EFsym1}  with \leqn{newfg}{
g(V)  = ( -\delta^{i}_4\delta^{j}_4 \chi_{\Rb}\rho(\alpha),
0,\ldots,0)^T \,, }
and we define the Newtonian potential by
\leqn{Npot}{
\Delta\Phi =  \chi_{\Rb}\rho \, .
}
Next, we use the Newtonian potential to define a new
combined gravitational-matter variable $W$ via the formula
\leqn{Wdef}{ W = V - d \Phi,} where \leqn{dPhi}{ d\Phi =
(0,\delta^i_4\delta^j_4\del_J\Phi(\alpha),0,0,0)\, .}
Notice that the transformation \eqref{Wdef} leaves the matter variables
unaffected. Consequently,  we can define $W$ by \eqn{Wdefa}{ W =
(\uf_4^{ij},W_I^{ij},\delta\uf^{ij},\alpha,w^i)^T, } and treat
$\Phi$ and $d\Phi$ as a function of $W$.
To formulate the
evolution equation entirely in terms of $W$, we need the
time derivative of the $\Phi$ map. So we define \lalign{Phid.1}{
\Phid(W,\ep U,\ep W, \ep^2 U) =
\Delta^{-1}\Bigl(\frac{2n \chi_{\Rb}\alpha^{2n-1}}{(4Kn(n+1))^n}\Pi&\bigl(a^4(\ep\uf,\ep\wv)^{-1}
\bigl[a^I(\wv,\ep\uf,\ep\wv)\del_I \wv \notag \\
& + b(\ufv,\wv,\ep\ufv,\ep\wv) \bigr] \bigr) \Bigr) \label{Phid}} where
$\Pi((\alpha,w^i)^T) = \alpha$ is a constant projection map.
By construction,  $\Phid = \partial_t\Phi$ when evaluated on a solution of
\eqref{EFsym1}.
To fit with the above notation, we also define
\eqn{dPhid}{
d\Phid = (0,\delta^i_4\delta^j_4\del_I\Phid,0,0,0)^T \, .
}
Noting that \leqn{shift}{
b^0(\ep V,\ep^2 U) = b^0(\ep W,\ep^2 U) \AND b^I(V,\ep U,\ep V, \ep^2) =
b^I(W,\ep U,\ep W,\ep^2 U), }
we can write \eqref{EFsym1} as
\lalign{wsysdef.1}{
b^0(\ep W,\ep^2 U)\del_t W = &
\frac{1}{\ep}c^I\del_I W + b^I(W,\ep U,\ep W,\ep^2 U)\del_I W \notag \\
&+
\Fc_0(W,\ep U,\ep W,\ep^2 U) + \ep \Fc_1(W,\ep U,\ep W,\ep^2 U),
\label{wsysdef}}
where
\lalign{Fcaldef}{
\Fc_0(W,&\ep U,\ep W,\ep^2 U) = f_0(W+d\Phi(W),\ep U,\ep(W+d\Phi(W)),\ep^2 U) \notag \\
& -b^0(\ep W,\ep^2 W)d\Phid(W,\ep U,\ep W,\ep^2 U)+ b^I(W,\ep U,\ep W)\del_I d\Phi(W),
\label{Fcaldef.1}
 \intertext{and}
\Fc_1(W,&\ep U,\ep W,\ep^2 U)) = f_1(W+d\Phi(W),\ep U,\ep(W+d\Phi(W)),\ep^2 U). \label{Fcaldef.2}
}

In the following Proposition, the constant $\Csob$ is defined to be the
$\ep$-independent in the weighted Sobolev inequality
$\norm{\cdot}_{W^{1,\infty}_{\eta,\ep}} \leq \Csob \norm{\cdot}_{H^\ell_{\eta,\ep}}$ which holds for $\ell > 3/2+1$ and
$0\leq \ep \leq \ep_0$ (see Lemma A.7 in \cite{Oli06} for a proof).
\begin{prop} \label{eplimA} \mnote{[eplimA]}
Suppose $-1 < \delta < -1/2$, $\ep_0>0$, $s\in \mathbb{N}_{0}$, $R>0$,
$K_1 < K_0/(2\sqrt{\ep_0}\Csob)$,  $\tau \geq 2K_1/\Csob$,
$\Rb > 16\tau+ R$,  $k\geq 3+s$, $\alphao,\wo^I \in H^k_{\delta-1}$,
$\supp\, \alphao \subset B_R$, $\mathfrak{z}^{IJ}\in H^{k+1}_\delta$,
$\mathfrak{z}^{IJ}_4 \in H^k_{\delta-1}$. Let
$\ufbo_\ep^{ij}$, $\del_t\ufbo^{ij}_\ep$ and $\wo^4_\ep$ be the
initial data constructed in Proposition \ref{idatA}, which,
by choosing $\ep_0\leq 1$ small enough, satisfies
\eqn{eplimA.1}{
\Bigl\|\Bigl(\ep\del_t\ufbo^{ij}_\ep,\del_I\ufbo^{ij}_\ep-\delta^i_4
\delta^j_4\del_I\Delta^{-1} \rhoo,0,\alphao,\wo^i_\ep\Bigr)^T\Bigr\|_{H^k_{\delta-1,\ep}} \leq K_1\, ,
\AND
\norm{\ufbo^{ij}_\ep}_{H^{k+1}_{\delta}}\leq \frac{K_0}{\sqrt{\ep_0}\Csob}
}for all $\ep \in (0,\ep_0]$.
Then there exists a $T>0$ independent of $\ep \in (0,\ep_0]$,
and maps
\eqn{eplimA.2}{
W_\ep = \bigl(\uf^{ij}_{4,\ep},W^{ij}_{I,\ep},\delta\uf^{ij}_\ep,
\alpha_\ep,w^i_\ep \bigr)^T \in X_{T_\ep,s,k,\delta-1} \qquad 0<\ep \leq \ep_0
}
such that
\begin{itemize}
\item[(i)] $T_\ep > T$ for $0\leq \ep \leq \ep_0$,
\item[(ii)] $W_\ep$ is the unique solution to
\eqref{wsysdef} with initial data
\eqn{eplimA.3}{
W_\ep(0) = \Bigl(\ep\del_t\ufbo^{ij}_\ep,\del_I\ufbo^{ij}-\delta^i_4
\delta^j_4\del_I\Delta^{-1} \rhoo,0,\alphao,\wo^i_\ep\Bigr)^T,
}
\item[(iii)]
\gath{epliA.5}{
\norm{W_\ep(t)}_{H^{k}_{\delta-1,\ep}} \leq 2K_1,
\quad
\ep \norm{\del_t W_{\ep}(t)}_{H^{k-1}_{\delta-1,\ep}} \lesssim 1,
\intertext{and}
\max\{\norm{\ep\ufb^{ij}_\ep(t)}_{L^\infty},
\norm{\ep\alpha_\ep(t)}_{L^\infty},\norm{\ep w^i(t)}_{L^\infty}\} < 2K_0
}
for all $(t,\ep) \in [0,T]\times (0,\ep_0]$,
\item[(iv)] for each $\ep \in (0,\ep_0]$, if
\gath{eplimA.5a}{
\limsup_{t\nearrow T_\ep}\norm{W_\ep(t)}_{W^{1,\infty}} <\infty \, ,
\intertext{and}
\sup_{0\leq t < T_\ep}\{\norm{\ep\ufb^{ij}_\ep(t)}_{L^\infty},
\norm{\ep\alpha_\ep(t)}_{L^\infty},\norm{\ep w^i(t)}_{L^\infty}\} < 2K_0 \, ,
}
then the solution $W_\ep(t)$ can be uniquely extended for some time $T^*_\ep>T_\ep$,

\item[(v)]
for any time $\tilde{T}_\ep$ which is strictly less than
the maximal existence time and for which
\eqn{eplimA.4a}{
\sup_{0\leq t \leq T_\ep}\{\norm{\ep\ufb^{ij}_\ep(t)}_{L^\infty},
\norm{\ep\alpha_\ep(t)}_{L^\infty},\norm{\ep w^i(t)}_{L^\infty}\} < 2K_0}
holds,
the support of $\alpha_\ep$ satisfies
\eqn{eplimA.4ab}{
\supp\, \alpha_\ep(t) \subset B_{\Rb_\ep} \quad \forall\; t \in [0,\tilde{T}_\ep]
}
where $\Rb_\ep := 16\sup_{0\leq t\leq \tilde{T}_\ep}\norm{w^I_\ep(t)}_{L^\infty}+R$,
\item[(vi)]$\supp\, \alpha_\ep(t) \subset B_{\Rb}$ for all
$(t,\ep) \in [0,T]\times (0,\ep_0]$,
\item[(vii)]$\del_t \ufb^{ij}_{\ep} = \ep^{-1} \ufb^{ij}_{4,\ep}$,
 and $\del_I \ufb^{ij}_{\ep} = W_{I,\ep}^{ij} +
\delta^{i}_4\delta^{j}_4\del_I\Phi(\alpha_\ep)$
where $\ufb^{ij}_\ep = \ufbo^{ij}_\ep + \ep^{-1}\delta \uf^{ij}$,
\item[(viii)] the triple $\{\ufb^{ij}_\ep,\alpha_\ep,w^i_\ep\}$ determines,
via the formulas \eqref{wdef.intro}, \eqref{dendef}, \eqref{den2met},
and \eqref{udensdef},
a solution to the full Einstein-Euler system \eqref{EEeqn} in
the harmonic gauge \eqref{harmB}
on the spacetime region $D_\ep = \Rbb^3\times [0,T]$, and
\item[(ix)] the conclusions {\rm (vii)-(viii)} continue to hold
on any region of the form $D_\ep = \Rbb^3\times [0,\tilde{T}_\ep]$ provided
$\supp\, \alpha_\ep(t) \subset B_{\Rb}$ for all $0\leq t\leq \tilde{T}_\ep$.
\end{itemize}
\end{prop}
\begin{proof}
\textbf{ (i)-(vii):} First we observe, that proof of statements (i)-(vii) follow from
a slight modification of the proof of Proposition 3.4 in \cite{Oli08}.

\bigskip

\noindent \textbf{(viii)-(ix):} Let $\psi_\ep$ satisfy the initial value problem
\leqn{eplimA10}{
\vb^k\delb_k\psi_\ep = 0 \quad : \quad \psi_\ep(0) = \chi_{3\Rb/2}(0),
}
and define
\eqn{eplimA11}{
\tilde{\Nc}_\ep = \psi\Nc_\ep = \psi\left(\ep \vb_i\vb^i + \frac{1}{\ep}\right).
}
Next, we observe that \eqref{eul10a.2} contracted with $\vb^i$ yields
\eqn{eplimA11}{
(1-2\ep\Nc_\ep)\vb^j\delb_j\Nc_\ep - \frac{\ep^2\alpha}{2nh}\vb^j\delb_j\alpha \Nc_\ep
+ \ep (1-2\ep\Nc_\ep)(\chi_{4\Rb}-1)\vb^i\vb^j \Gammab_{ij}^k \vb_k = 0.
}
Multiplying this equation by $\psi_\ep$ then gives
\leqn{eplimA12}{
(1-2\ep\Nc_\ep)vb^j\delb_j\tilde{\Nc}_\ep - \frac{\ep^2\alpha}{2nh}\vb^j\delb_j\alpha \tilde{\Nc}_\ep
+ \ep(1-2\ep\Nc_\ep)\psi_\ep(\chi_{4\Rb}-1)\vb^i\vb^j \Gammab_{ij}^k \vb_k = 0.
}

From statements (i)-(vii), we have that
\leqn{eplimA13}{ w^i_\ep(t) \in X_{T,s,k,\delta-1}
}
and
\leqn{eplimA14}{
\norm{w^i(t)}_{H^k} \lesssim \norm{w^i(t)}_{H^k_{\delta-1,\ep}} \lesssim 1
\qquad \forall \;(t,\ep) \in [0,T)\times (0,\ep].
}
Since
\eqn{eplimA15}{
\vb^j\del_i = (1+\ep w^4_\ep)\del_t + w^I_\ep\del_I,
}
we get from \eqref{eplimA13}, \eqref{eplimA14}, and the
hyperbolic equation \eqref{eplimA10} that
$\psi \in X_{T,s,k,\delta-1}$ and
\eqn{eplimA16}{
\norm{\psi_\ep(t)}_{H^k}\lesssim 1 \qquad \forall \; (t,\ep) \in [0,T)\times (0,\ep].
}
From the finite propagation speed property of hyperbolic equations, we conclude that
there exists a time $T_*\in (0,T)$ such that
\leqn{eplimA17}{
\psi_\ep(t)|_{B_{\Rb}} = 1, \AND \supp \psi_\ep(t) \subset B_{4\Rb} \qquad \forall \; (t,\ep)\in [0,T_*)
\times (0,\ep_0].
}
In particular, this implies that
\leqn{eplimA18}{
\Nc_\ep(t,x) = \tilde{\Nc}_\ep (t,x) \qquad \forall \; (t,x,\ep) \in [0,T_*)\times B_{\Rb} \times (0,\ep_0],
}
and
\leqn{eplimA19}{
(\chi_{4\Rb}(x)-1)\psi_\ep(t,x) = 0 \qquad  \forall \; (t,x,\ep) \in [0,T_*)\times \Rbb^3 \times (0,\ep_0].
}
Using \eqref{eplimA19}, equation \eqref{eplimA12} reduces to
\eqn{eplimA20}{
(1-2\ep\Nc_\ep)vb^j\delb_j\tilde{\Nc}_\ep - \frac{\ep^2\alpha}{2nh}\vb^j\delb_j\alpha \tilde{\Nc}_\ep =0
}
for all $(t,x,\ep) \in [0,T_*)\times \Rbb^3 \times (0,\ep_0]$. But
$\Nc_\ep(0) = 0$ from the choice of initial data which implies that $\tilde{\Nc}_\ep(0)=0$.
By the uniqueness of solutions to hyperbolic equations, we conclude that
$\tilde{\Nc}_{\ep}(t,x) = 0$ for all $(t,x,\ep) \in [0,T_*)\times \Rbb^3 \times (0,\ep_0]$, and
hence $\Nc_\ep(t,x) = 0$ for all $(t,x,\ep) \in [0,T_*)\times B_\Rbb \times (0,\ep_0]$. This
implies that the fluid velocity normalization $\vb^i \vb_i = -1/\ep^2$ is
satisfied for all  $(t,x,\ep) \in [0,T_*)\times B_\Rbb \times (0,\ep_0]$. Using this
and the fact that $\supp\alpha_\ep(t) \subset B_\Rbb$ for all $(t,\ep)\in [0,T_*)\times (0,\ep_0]$,
it is not difficult to verify from the evolution equation \eqref{eul10a.1}-\eqref{eul10a.2}
that
\eqn{eplimA21}{
\{\vb^4(t,x)= 1+\ep w^4_\ep(t,x), \vb^I(t,x)=w^I_\ep(t,x),
\rho_\ep(t,x) := (4Kn(n+1))^{-n}\alpha_\ep^{2n}(t,x)\}
}
satisfy
the Euler equations \eqref{eul6} (or equivalently \eqref{eul1}) for all
$(t,x,\ep) \in [0,T_*)\times \Rbb^3 \times (0,\ep_0]$. With the Euler equations satisfied,
the remainder of the proof follows as in the proof of Proposition 3.4 in \cite{Oli08}.
\end{proof}

%% file: limiteqns.tex
\sect{limiteqns}{Limit equations}

In this section, we describe the limit equations that
govern the gravitational and matter fields in
the limit $\ep\searrow 0$. We show in the
next section that solutions to these equations
approximate the solutions to the
full Einstein-Euler equations up to a remainder
term that is of order $\ep$.

\subsect{flim}{Fluid limit equations}

The fluid limit equations are
\lalign{newtB}{
\del_t\alphat &= -\wt^I\del_I
\alphat -\frac{\alphat}{2n}\del_I\wt^I, \label{newtB.1} \\
\del_t \wt^J &= -\frac{\alphat}{2n}\del^J\alphat - \wt^I\del_I
\wt^J -\chi_{4\Rb}\del^J\Phit \label{newtB.2}, \\
\Delta \Phit &= \rhot \qquad \bigl(\rhot :=
(4Kn(n+1))^{-n}\alphat^{2n}\bigr). \label{newtB.3} }

\begin{prop} \label{cogA} \mnote{[cogA]}
Let $k$, $s$, $\Rb$, $\delta$, $\alphao$, and $\wo$ be as in Proposition \ref{eplimA}.
Then there exists a maximal time $T_0^M>0$ and a unique solution \gath{cogA1}{ \alphat, \wt^I
\in C^0([0,T^M_0),H^k_{\delta-1})\cap
C^{1}([0,T_0),H^{k-1}_{\delta-1})\, , \\
\Phit \in C^0([0,T^M_0),H^{k+2}_{\delta})\cap
C^{1}([0,T^M_0),H^{k+1}_{\delta})\, , \quad \del_t\Phit \in
C^{0}([0,T^M_0),H^{k+1}_{\delta-1}) } to
\eqref{newtB.1}-\eqref{newtB.3} satisfying $\alphat(0)=\alphao$ and  $\wt^I(0)=\wo^I$. Moreover,
\eqn{cogA2a}{\alphat,
\wt^I \in X_{T^M_0,s,k,\delta-1} \, , \quad
\Phit \in X_{T^M_0,s,k+2,\delta} \, , \quad \del_t\Phit =-\del_I\Delta^{-1}
(\rhot \wt^I) \in X_{T^M_0,s,k+1,\delta-1} \, , }
and
\eqn{cogA2}{
\supp\, \alphat(t) \subset B_{R(t)}
}
where $R(t)=R+t\sup_{0\leq s\leq t}\norm{\wt^I(s)}_{L^\infty}$.
\end{prop}
\begin{proof}
The proof follows from a trivial modification of the proof in Proposition 3.7 in \cite{Oli08}.
\end{proof}

\begin{rem} \label{cogB} \mnote{[cogB]}
Since $\Rb > R$, it is clear from Proposition \ref{cogA} and the weighted Sobolev inequality (see Lemma A.7 in \cite{Oli06})
that there exists a time $T_0 \in (0,T_M)$ such that
\leqn{cogB1}{
\supp\, \alphat(t) \subset B_{4\Rb} \quad \text{for all $t\in [0,T_0]$.}
}
In particular, this shows that
\leqn{cogB2}{
\rhot(t)\chi_{4\Rb} = \rhot(t) \quad \text{for all $t\in [0,T_0]$,}
}
which in turn implies that the pair $\{\rhot(t),\wt^I(t)\}$ satisfies
the Poisson-Euler equations \eqref{newtPE.1}-\eqref{newtPE.3} on the
time interval $[0,T_0]$.
\end{rem}

\subsect{glim}{Gravitational limit equations}

The gravitational limit equations are defined
by
\leqn{glim1}{
\del_t \Xve = \frac{1}{\ep} C^I\del_I \Xve + (0,0,X_{4,\ep}^{ij})^T,
}
where
\eqn{glim2}{
\Xve = (X_{4,\ep}^{ij}, X_{I,\ep}^{ij},X^{ij}_\ep)^T .
}

\begin{prop} \label{glimA} \mnote{[glimA]}
Let $\delta$, $k$, $\alphao$, $\del_t\ufbo{}_\ep^{ij}$, $\del_I\ufbo{}_\ep^{ij}$
be as in  Proposition \ref{eplimA}, and
\leqn{glimA1}{
\Xve(0) = \Bigl(\ep\del_t\ufbo{}^{ij}_{\ep},\del_I\ufbo{}_\ep^{ij}-\del_I\Delta^{-1}(\delta^i_4\delta^j_4\rhoo),0
\Bigr)^T\Bigl|_{\ep=0}.
}
Then there exists a unique solution
\eqn{glimA2}{
\Xve \in C^0([0,\infty),H^{k}_{\delta-1})\cap C^{1}([0,\infty),H^{k-1}_{\delta-1})
}
to \eqref{glim1} with initial data \eqref{glimA1} that satisfies $\Xve \in X_{\infty,s,k,\delta-1}$
and the estimates:
\begin{itemize}
\item[(i)]\alin{glimA3}{
&\norm{\Xve(t)}_{H^k_{\delta-1,\ep}}+ \ep\norm{\del_t\Xve(t)}_{H^{k-1}_{\delta-1,\ep}} \lesssim e^{Ct}, \\
&\norm{X^{ij}_\ep(t)}_{L^\infty_{\delta,\ep}} + \norm{DX^{ij}_\ep(t)}_{H^{k-1}_{\delta-1,\ep}} \lesssim
e^{Ct}\ep,
}
for all $(t,\ep)\in [0,\infty)\times (0,\ep_0]$ and some fixed constant $C>0$, and
\item[(ii)] for
any $\Lambda >0$,
\eqn{glimA3a}{
\norm{X^{ij}_{I,\ep}(t)}_{W^{\ell,\infty}(B_\Lambda(\Rbb^3)} + \norm{X^{ij}_{4,\ep}(t)}_{W^{\ell,\infty}(B_\Lambda(\Rbb^3))} \lesssim
\frac{\ep^{3/2}\sqrt{\Lambda+1}}{(\ep+t)^{3/2}}
\qquad  0\leq \ell < k-3/2
}
for all $(t,\ep)\in [0,\infty)\times (0,\ep_0]$.
\end{itemize}
\end{prop}
\begin{proof}
Since $-1<\delta<-1/2$,
it follows from Lemma A.11 of \cite{Oli06} and Proposition \ref{idatA} that
\leqn{glimA4}{
\norm{\Xve(0)}_{H^{k}_{\delta-1},\ep} \lesssim \norm{\Xve(0)}_{H^{k}_{\delta-1}} \lesssim 1.
}
This inequality together with the weighted energy estimates (see Lemma 7.1 in \cite{Oli06})
gives
\leqn{glimA5}{
\norm{\Xve(t)}_{H^k_{\delta-1}} \lesssim e^{C t}\norm{\Xve(0)}_{H^k_{\delta-1,\ep}}\lesssim e^{C t}.
}
for some fixed positive constant $C$.

From the evolution equation \eqref{glim1} and the choice of initial data, we see that
\eqn{glimA6}{
\del_t\bigl( \del_{I}X_{J,\ep}^{ij}(t)-\del_{J}X^{ij}_{I,\ep}\bigr) = 0 \AND \del_I X_{J,\ep}^{ij}(0)-\del_J X_{I,\ep}^{ij} = 0,
}
which implies that
\leqn{glimA7}{
\del_{I}X_{J,\ep}^{ij}(t)- \del_{I}X_{J,\ep}^{ij}(t)=0.
}
Also, it is not difficult to show that
\leqn{glimA8}{
\del_t\bigl(\del_I X^{ij}_\ep - \ep X^{ij}_{I,\ep}) = 0
}
follows from \eqref{glim1}. Combining \eqref{glimA7} and \eqref{glimA8} then
yields
\leqn{glimA9}{
\del_I X^{ij}_\ep(t) = \ep\bigl( X^{ij}_{I,\ep}(t) - X^{ij}_{I,\ep}(0) \bigr).
}
Next, we note that
\leqn{glimA10}{
\norm{X^{ij}_\ep}_{L^\infty_{\delta,\ep}} \lesssim \norm{DX^{ij}_\ep}_{H^k_{\delta-1,\ep}} + \ep
\norm{X^{ij}_\ep}_{L^2_{\delta-1,\ep}}
}
follows from the weighted Sobolev inequalities (see Lemma A.7 in \cite{Oli06}). Collecting the estimates
\eqref{glimA5}, \eqref{glimA9}, and \eqref{glimA10}, we arrive at
\eqn{glimA12}{
\norm{X^{ij}_\ep(t)}_{L^\infty_{\delta,\ep}}
+ \norm{D X^{ij}_\ep}_{H^k_{\delta-1,\ep}} \lesssim e^{Ct}\ep.
}

To prove the last two estimates for  $X^{ij}_{I,\ep}$ and $X^{4}_{I,\ep}$ , we observe that $X^{ij}_{I,\ep}$ and $X^{4}_{I,\ep}$ satisfy
the wave equations
\leqn{glimA12}{
\ep^2 \del_t^2 X^{ij}_{I,\ep} - \Delta X^{ij}_{I,\ep}=0  \AND \ep^2 \del_t^2 X^{ij}_{4,\ep}-\Delta
X^{ij}_{I,\ep} = 0.
}
Since the initial data for these equations satisfy \eqref{glimA4} and $-1<\delta < -1/2$, we can
apply the weighted dispersive estimates from Theorem 1.1 in \cite{DGK01} to obtain
\lalign{glimA13}{
\bigl| D_x^\ell X^{ij}_{I,\ep}(t,x)\bigr| &\lesssim
\frac{\norm{X^{ij}_{I,\ep}(0)}_{H^k_{\delta-1}} + \norm{\del_I X^{ij}_{4,\ep}(0)}_{H^{k-1}_{\delta-2}}}{(1+t/\ep+|x|)\sqrt{|1+|t/\ep-|x||}},
\label{glimA13.1}
\intertext{and}
\bigl| D_x^\ell X^{ij}_{4,\ep}(t,x)\bigr| &\lesssim \frac{\norm{X^{ij}_{4,\ep}(0)}_{H^k_{\delta-1}} + \norm{\del^I X^{ij}_{I,\ep}(0)}_{H^{k-1}_{\delta-2}}}{(1+t/\ep+|x|)\sqrt{|1+|t/\ep-|x||}}
\label{glimA13.2}}
for $0\leq \ell < k-3/2$. But
\eqn{glimA14}{
1+|t/\ep| \leq 1 + |t/\ep -|x|+|x|| \leq 1+\Lambda + |t/\ep-|x||
\leq (\Lambda +1)(1+|t/\ep-|x||)}
for $|x|\leq \Lambda$, and so
the inequalities \eqref{glimA13.1} and \eqref{glimA13.2} imply that
\eqn{glimA13}{
\norm{X^{ij}_{I,\ep}(t)}_{W^{\ell,\infty}(B_\Lambda(\Rbb^3))} +
\norm{ X^{ij}_{4,\ep}(t)}_{W^{\ell,\infty}(B_\Lambda(\Rbb^3))} \lesssim
\frac{\ep^{3/2}\sqrt{\Lambda+1}}{(\ep+t)^{3/2}}
}
for $0\leq \ell < k-3/2$.
\end{proof}

\begin{rem} \label{glimB} \mnote{[glimB]} From the initial value problem \eqref{glim1}-\eqref{glimA1} (see \eqref{glimA9})
and Propositions \ref{idatA} and \ref{cogA}, it is not difficult to verify that
\leqn{glimB1}{
\ufbt_\ep^{ij} := \frac{1}{\ep} X^{ij}_\ep + \bigl(\delta^i_I\delta^j_J\zf^{IJ}
-2\Delta^{-1}\del_I\zf_4^{IJ}\delta^{(i}_4\delta^{j)}_J + \delta^{i}_4\delta^j_4 \Delta^{-1}(\rhoo+\del^2_{IJ}\zf^{IJ})\bigr)
+(\Phit(t)-\Phit(0))\delta^{i}_4\delta^j_4
}
satisfies the identities
\leqn{glimB4}{
\del_t\ufbt^{ij}_{\ep} = \frac{1}{\ep} X^{ij}_{4,\ep} + \delta_4^i\delta_4^j\del_t\Phit, \quad
\del_I\ufbt^{ij}_\ep = X^{ij}_{I,\ep} + \delta_4^i\delta_4^j\del_I \Phit,
}
and the wave equation
\leqn{glimB2}{
\ep^2\del_t^2 \ufbt_\ep^{ij} - \Delta \ufbt_\ep^{ij} = -\delta^i_4\delta^j_4 \rhot
+\delta^i_4\delta^j_4\ep^2\del_t^2\Phit,
}
with initial conditions
\lalign{glimB3}{
\ufbt^{ij}_\ep\bigl|_{t=0} & = \delta^i_I\delta^j_J\zf^{IJ}
-2\Delta^{-1}\del_I\zf_4^{IJ}\delta^{(i}_4\delta^{j)}_J + \delta^{i}_4\delta^j_4
(\Phit\bigl|_{t=0}+\Delta^{-1}\del^2_{IJ}\zf^{IJ}), \label{glimB3.1}\\
\del_t\ufbt^{ij}_\ep\bigl|_{t=0} & = \frac{1}{\ep} \bigl(\delta^i_I\delta^j_J\zf_4^{IJ}
-2\del_I\zf^{IJ}\delta^{(i}_4\delta^{j)}_J + \delta^{i}_4\delta^j_4
\Delta^{-1} \del^2_{IJ}\zf_4^{IJ} \bigr) + \delta_4^i\delta_4^j\del_t\Phit\bigl|_{t=0} . \label{glimB3.2}
}
\end{rem}

\subsect{clim}{The combined system}

Collecting the fluid and gravitation limit variables into a single vector
\leqn{Ydef}{
\Ye = \bigl(\Xve,\alphat,\wt^I,\wt^4\bigr)^T,
}
we can write the combined gravitational-fluid limit equations in
the following form:
\leqn{cseqns1}{
\del_t \Ye = \frac{1}{\ep}c^I\del_I Y + \bt_0^I\del_I\Ye + \Fct_0+ c^I\del_I\omega
}
where
\leqn{cseqns2}{
\bt_0^I = \begin{pmatrix} 0 & 0 \\
0 & \at^I \end{pmatrix}, \qquad \at^I = \begin{pmatrix} -\wt^I & -\frac{\alphat}{2n}\delta^I_j \\
-\frac{\alphat}{2n}\delta^I_i & -\delta_{ij} \wt^I \end{pmatrix},
}
\leqn{cseqns3}{
\omega = (\omega_4^{ij}, \omega_I^{ij},0,0,0,0)^T, \quad \omega_4^{ij} = \del_t\Phit\delta^i_4\delta^j_4,
\quad \omega^{ij}_I =\del_I \Delta^{-1}\big(2\rho \wt^J \delta_J^{(i}\delta_4^{j)} \bigr),
}
and
\leqn{cseqns4}{
\Fct_0 =
\bigr(-2\rho \wt^J \delta_J^{(i}\delta_4^{j)},-\del_I\del_t\Phit\delta^i_4\delta^4_j,X^{ij}_4,0,-\chi_{4\Rb}\del^I\Phit,0\bigl)^T.
}

%% file: nlim.tex
\sect{nlim}{The fast Newtonian limit}

We begin by defining the error $\Ze$ between the limit $\Ye$ and the full solution
$\We$ by
\leqn{Zdef}{
\We = \Ye + \ep(\omega_\ep + \Ze).
}
Next, we let
\leqn{Zeqn1}{
\Fct = \Fc_0(Y_\ep,0,0,0), \AND \bt^I = b^I(Y_\ep,0,0,0),
}
and observe that
\leqn{Zeqn2}{
\Fct = \Fct_0 + \ep \Fct_1, \AND \bt^I = \bt_0^I + \ep \bt_1^I,
}
where
\leqn{Zeqn3}{
\bt_1^I = \frac{1}{\ep}\begin{pmatrix}\At^I & 0 \\ 0 & 0 \end{pmatrix},
\quad
\At^I = \begin{pmatrix}
8 X^{4I}_\ep & 4X^{IJ}_\ep & 0 \\
4X^{IJ}_\ep & 0 &  0\\
0 & 0 & 0
\end{pmatrix},
}
and
\leqn{Zeqn4}{
\Fct_1 = \frac{1}{\ep} \Bigl(4\bigl(
X^{IJ}_\ep\del^2_{IJ}\Phit -\rhot\eta_{ij}X^{ij}_\ep \bigr) \delta^{i}_4\delta^{j}_4,0,0,
0,\Fct_1^i \Bigr)^T.
}
with
\eqn{Zeqn4a}{
\Fct_1^I = -\chi_{4\Rb}\bigl[\delta^{IJ}\bigl(X^{44}_{J,\ep}+\delta_{KL}X^{KL}_{I,\ep}\bigr)+4X^{J4}_4\bigr],
}
and
\eqn{Zeqn4b}{
\Fct_1^4 =
-\chi_{4\Rb}\bigl(X^{44}_{4,\ep}+\delta_{KL}X^{KL}_{4,\ep} \bigr).
}

Now, let $T_* = \min\{T_0,T\}$ where $T$ and $T_0$ are as defined in Propositions \ref{eplimA} and Remark \ref{cogB}, respectively. Then by Propositions \ref{eplimA}, \ref{cogA}, and \ref{glim},
for any $\ep \in (0,\ep_0]$, the error $\Ze(t)$ (see  \eqref{wsysdef}, \eqref{cseqns1}, and
\eqref{Zdef}-\eqref{Zeqn2}) satisfies the
initial value problem
\lalign{ZeqnA}{
b^0_\ep \del_t \Ze &= \frac{1}{\ep}c^I\del_I \Ze + b^I_\ep \del_I \Ze +
\Rc_\ep \label{ZeqnA.1} \\
\Ze(0) &=  \frac{1}{\ep}\bigl(W_\ep(0)-Y_\ep(0))-\omega(0) ,\label{ZeqnA.2}
}
on the interval $0\leq t\leq T_*$,
where
\lgath{ZeqnB}{
b^0_\ep = b^0(\ep \We,\ep^2 \Ue), \quad b^I_\ep = b^I(\We,\ep \Ue,\ep \We,\ep^2 \Ue),
\label{ZeqnB.1} \\
\Fc_\ep = \Fc_0(\We,\ep \Ue,\ep \We,\ep^2 \Ue) + \ep \Fc_1(\We,\ep \Ue,\ep \We,\ep^2 \Ue),
\label{ZeqnB.2}
\intertext{and}
\Rc_\ep = b^I_\ep\del_I\omega-b^0_\ep\del_t\omega+ \frac{b^I_\ep-\bt^I}{\ep}\del_I Y + \frac{\Fc_\ep-\Fct}{\ep}
+\frac{b^0_\ep-\id}{\ep^2}\ep\del_t Y + \bt_1^I\del_I Y + \Fct_1\, .
}

\begin{prop} \label{nlimA} \mnote{[nlimA]}
Let $\delta$, $k$, $s\geq 2$, $T$, and $W_\ep(t)$ be as in Proposition \ref{eplimA}, $T_0$ as in Remark \ref{cogB},
$Y_\ep(t)$ as defined by \eqref{Ydef}, and $T_*=\min\{T_0,T\}$.
Then for $\ep_0>0$ small enough
\eqn{nlimA1}{
\norm{W_\ep(t)-Y_\ep(t)}_{H^{k-2}_{\delta-1,\ep}} \lesssim \ep
}
for all $(t,\ep)\in [0,T_*)\times (0,\ep_0]$.
\end{prop}
\begin{proof} By Propositions \ref{idatA}, \ref{eplimA}, and \ref{cogA}, there exists a positive constant $C_0$ such that
\leqn{nlimA6}{
\norm{Z_\ep(0)}_{H^k_{\delta-1}}\leq C_0 \quad \text{for all $\ep \in (0,\ep_0]$}.
}
Next, choosing $\ep_0$ small enough, it follows directly from Propositions 3.5 and 3.6 of \cite{Oli08} and Propositions \ref{cogA} and \ref{glimA} of the previous section that
\leqn{nlimA7}{
\Bigl\|b^I_\ep\del_I\omega - b^0_\ep\del_t \omega +\frac{b^I_\ep-\bt^I}{\ep}\del_I Y_\ep
+ \frac{\Fc_\ep-\Ft_0}{\ep} \Bigr\|_{H^{k-2}_{\delta-1,\ep}} \lesssim 1 + \norm{Z_\ep}_{H^{k-2}_{\delta-1,\ep}}
}
for all $(t,\ep)\in [0,T_*)\times (0,\ep_0]$ provided $\norm{Z_\ep(t)}_{H^{k-2}_{\delta-1,\ep}}\leq 2C_0/\ep$.
Also, from Lemmas A.1 and A.4 of \cite{Oli06}, and
Proposition \ref{cogA} and \ref{glimA}, we see that
\leqn{nlimA8}{
\Bigl\|\frac{b^0_\ep-\id}{\ep^2}\ep\del_t Y_\ep + \bt^I_1\del_I Y_\ep + \Fct_1 \Bigr\|_{H^{k-2}_{\delta-1,\ep}} \lesssim 1 +
\frac{\sqrt{\ep}}{(\ep+t)^{3/2}}
}
for all $(t,\ep)\in  [0,T_*)\times (0,\ep_0]$.

Defining the energy norm
\eqn{nlimA8a}{
\nnorm{\cdot}_{{k-2},\delta,\ep} := \sum_{|\alpha|\leq {k-2}}
\ip{D_x^\alpha(\cdot)}{b^0_\ep D_x^\alpha (\cdot)} \, ,}
we see via Proposition \ref{eplimA}
that
\eqn{nlimA9}{
\norm{\cdot}_{H^{k-2}_{\delta-1,\ep}} \lesssim
\nnorm{\cdot}_{{k-2},\delta-1,\ep} \lesssim
\norm{\cdot}_{H^{k-2}_{\delta-1,\ep}},
}
uniformly for $(t,\ep)\in  [0,T_*)\times (0,\ep_0]$. Setting $\Zc_\ep = \ep Z_\ep$, the
evolution equation \eqref{ZeqnA.1} and  the weighted energy estimates (see the proof
of Theorem B.1 in \cite{Oli08}) in conjunction
with Proposition \ref{eplimA} and the  estimates \eqref{nlimA7}-\eqref{nlimA8} show
that there exists a fixed constant $C_1>0$ such that
\eqn{nlimA9}{
\frac{d\;}{dt}\nnorm{\Zc_\ep(t)}_{{k-2},\delta-1,\ep} \leq C_1\left( \nnorm{\Zc_\ep(t)}_{{k-2},\delta-1,\ep} +
 \ep+ \left(\frac{\ep}{\ep+t}\right)^{3/2} \right),
}
for all $t$ such that $\norm{\Zc_\ep(t)}_{{k-2},\delta-1,\ep} \leq 2C_0$.
Gronwall's inequality and \eqref{nlimA6} then show that
\alin{nlimA10}{
\nnorm{\Zc_\ep(t)}_{{k-2},\delta-1,\ep}& \leq  e^{C_1 t} C_0\ep + \ep\int_{0}^t e^{C_1(t-s)}\, ds +
\ep^{3/2}\int_{0}^t \frac{e^{C_1(t-s)}}{(\ep+s)^{3/2}} \, ds \\
& \leq \ep e^{C_1 t}\left( C_0 + 1 + \sqrt{\ep}\int_{0}^t \frac{1}{(\ep+s)^{3/2}}\, ds \right) \\
& \leq \ep e^{C_1 t}(C_0 + 3),
}
again for all $t$ such that $\norm{\Zc_\ep(t)}_{{k-2},\delta-1,\ep} \leq 2C_0$. Therefore choosing
$\ep_0>0$ small enough we obtain
\eqn{nlimA11}{
\norm{Z_\ep(t)}_{H^{k-2}_{\delta-1,\ep}} \lesssim 1
}
for all $(t,\ep)\in  [0,T_*)\times (0,\ep_0]$, and the proof is complete.
\end{proof}

We are now ready to prove the main theorem.

\begin{proof}[Proof of Theorem \ref{mthm}]
$\;$

\smallskip

\noindent \textbf{(i):} Since the ADM mass is conserved, statement (i) follows
directly from Corollary \ref{mass} and Proposition \ref{eplimA}.

\bigskip

\noindent \textbf{(ii)-(iv):} From the definition of $W_\ep$ and $Y_\ep$, we have
\eqn{mthm2}{
\norm{\alpha_\ep(t)-\alphat(t)}_{H^{k-2}_{\delta-1,\ep}}  + \norm{w^I_\ep(t)-\wt^I(t)}_{H^{k-2}_{\delta-1,\ep}} +
\norm{w^4_\ep(t)}_{H^{k-2}_{\delta-1,\ep}} \leq  \norm{W_\ep(t)-Y_\ep(t)}_{H^{k-2}_{\delta-1,\ep}},
}
and hence, by Proposition \ref{nlimA},
\leqn{mthm3}{
\norm{\alpha_\ep(t)-\alphat(t)}_{H^{k-2}_{\delta-1,\ep}}  + \norm{w^I_\ep(t)-\wt^I(t)}_{H^{k-2}_{\delta-1,\ep}} +
\norm{w^4_\ep(t)}_{H^{k-2}_{\delta-1,\ep}} \lesssim \ep
}
for all $(t,\ep)\in [0,T_*)\times (0,\ep_0]$. Also by the weighted multiplication Lemma (see Lemma
A.8 in \cite{Oli06}) and Propositions \ref{eplimA} and \ref{cogA}, we have
\leqn{mthm4}{
\norm{\rhot_\ep(t)-\rhot_\ep(t)}_{H^{k-2}_{\delta-1,\ep}} \lesssim \norm{\alpha_\ep(t)-\alphat(t)}_{H^{k-2}_{\delta-1,\ep}}
}
for all $t\in [0,T_*)\times (0,\ep_0]$,
while
\lalign{mthm1}{
\norm{\rho_\ep(t)-\rhot(t)}_{H^{k-2}} & + \norm{w^I_\ep(t)-\wt^I(t)}_{H^{k-2}} + \norm{w^4_\ep(t)}_{H^{k-2}}\notag \\
& \lesssim  \norm{\rho_\ep(t)-\rhot(t)}_{H^{k-2}_{\delta-1,\ep}} + \norm{w^I_\ep(t)-\wt^I(t)}_{H^{k-2}_{\delta-1,\ep}} + \norm{w^4_\ep(t)}_{H^{k-2}_{\delta-1,\ep}} \label{mthm1.1}
}
is a consequence of Lemma A.11 and equation (A.24) of \cite{Oli06}. Combining the inequalities
\eqref{mthm3}-\eqref{mthm1.1}, we arrive at
\lalign{mthm1}{
\norm{\rho_\ep(t)-\rhot(t)}_{H^{k-2}} & + \norm{w^I_\ep(t)-\wt^I(t)}_{H^{k-2}} + \norm{w^4_\ep(t)}_{H^{k-2}}\lesssim \ep
}
for all $t\in [0,T_*)\times (0,\ep_0]$.

Next, we observe that
\eqn{mathm5}{
\norm{\del_I\Phi_\ep(t)-\del_I\Phit(t)}_{H^{k-2}_{\delta-1,\ep}}
+\norm{\ep\del_t\Phi_\ep(t)}_{H^{k-2}_{\delta-1,\ep}}
\lesssim \ep
}
for all $(t,\ep) \in [0,T_*)\times (0,\ep_0]$ by Propositions \ref{eplimA} and \ref{nlimA}, and Lemmas 3.2 and 3.3 of \cite{Oli08}.
From the above estimate, the identities $W^{ij}_{I,\ep}=\del_I\ufb^{ij}_\ep+\delta^{i}_4\delta^j_4\del_I\Phi_\ep$
and $\uf_{4,\ep}^{ij} = \ep\del_t\ufb^{ij}_\ep$ (see
Proposition \ref{eplimA}),
and the relations \eqref{glimB1}-\eqref{glimB4}, we get
\alin{mathm6}{
\norm{\del_I\ufb^{ij}_\ep(t)-\del_I\ufbt^{ij}_\ep (t)}_{H^{k-2}_{\delta-1,\ep}}
&+\norm{\ep\del_t\ufb^{ij}_\ep(t)-\ep \del_t \ufbt^{ij}_\ep (t)}_{H^{k-2}_{\delta-1,\ep}} \\
\lesssim & \norm{W_{I,\ep}^{ij}(t)-X_{I,\ep}^{ij}(t)}_{H^{k-2}_{\delta-1,\ep}} +
 \norm{\uf_{4,\ep}^{ij}(t)-X_{4,\ep}^{ij}(t)}_{H^{k-2}_{\delta-1,\ep}} + \ep \\
\lesssim &  \norm{W_\ep(t)-Y_\ep(t)}_{H^{k-2}_{\delta-1,\ep}} + \ep
}
and hence, by  Proposition \ref{nlimA}
and Lemma A.7 of \cite{Oli06},
\eqn{mathm7}{
\norm{\ufb^{ij}_\ep(t)-\ufbt^{ij}_\ep(t)}_{L^6_{\delta,\ep}} + \norm{\del_I\ufb^{ij}_\ep(t)-\del_I\ufbt^{ij}_\ep (t)}_{H^{k-2}_{\delta-1,\ep}}
+\norm{\ep\del_t\ufb^{ij}_\ep(t)-\ep \del_t \ufbt^{ij}_\ep (t)}_{H^{k-2}_{\delta-1,\ep}} \lesssim \ep
}
for all $(t,\ep) \in [0,T_*)\times (0,\ep_0]$. Finally, it follows from Lemma A.11 and equation (A.24) of \cite{Oli06}, and the above estimate
that
\eqn{mathm7}{
\norm{\ufb^{ij}_\ep(t)-\ufbt^{ij}_\ep(t)}_{L^6} + \norm{\del_I\ufb^{ij}_\ep(t)-\del_I\ufbt^{ij}_\ep (t)}_{H^{k-2}}
+\norm{\ep\del_t\ufb^{ij}_\ep(t)-\ep \del_t \ufbt^{ij}_\ep (t)}_{H^{k-2}} \lesssim \ep
}
for all $(t,\ep) \in [0,T_*)\times (0,\ep_0]$. This completes the proof.
\end{proof}

%% file: fnlim_atmp.bbl
\begin{thebibliography}{10}

\bibitem{Bart05} R.~ Bartnik, {\em Phase Space for the Einstein Equations}, 
 Comm. Anal. Geom. \textbf{13} (2005), 845-885. 
\bibitem{BrKa07} U. Brauer and L. Karp, \emph{
Local existence of classical solutions for the Einstein-Euler system using weighted Sobolev spaces of fractional order}, 
 C. R. Acad. Sci. Paris, Ser. I {\bf 345} (2007), 49-54.
\bibitem{BK} G.~Browning and H.O.~Kreiss, {\em Problems with
different time scales for nonlinear partial differential
equations}, SIAM J. Appl. Math. {\bf 42} (1982), 704-718.
\bibitem{DGK01} P.~D'Ancona, V.~Georgiev, and H.~Kubo, {\em Weighted decay estimates
for the wave equation}, J. Differential Equations {\bf 177} (2001), 146-208.
\bibitem{Iso87} H.~Isozaki, {\em Wave operators and the incompressible limit of
the compressible Euler equations}, Comm. Math. Phys. {\bf 110} (1987), 519-524.
\bibitem{KM82} S.~Klainerman and A.~Majda, {\em Compressible and incompressible fluids},
Comm. Pure Appl. Math. {\bf 35} (1982), 629-651.
\bibitem{Kreiss} H.O.~ Kreiss, \emph{Problems with different time
scales for partial differential equations}, Comm. Pure Appl. Math.
{\bf 33} (1980), 399-439.
\bibitem{Lott} M.~Lottermoser, \emph{A convergent post-Newtonian approximation for the
constraints in general relativity}, Ann. Inst. Henri Poincar\'{e} \textbf{57} (1992), 279-317.
\bibitem{Mak} T.~Makino, ``On a local existence
theorem for the evolution equation of gaseous stars'',
in {\em Patterns and Waves}, edited by T.~Nishida,
M.~Mimura, and H.~Fujii, North-Holland, Amsterdam, 1986.
\bibitem{Oli06} T.A.~Oliynyk, \emph{The Newtonian limit for perfect fluids},
Comm. Math. Phys. \textbf{276} (2007), 131-188.
\bibitem{Oli08} T.A.~Oliynyk, \emph{Post-Newtonian expansions for perfect
fluids},
Comm. Math. Phys. (accepted).
\bibitem{Oli09} T.A.~Oliynyk, \emph{On the consistency of the post-Newtonian expansion in general relativity}, in preparation.
\bibitem{Ren92} A.D.~Rendall, {\em The initial value problem
for a class of general relativistic fluid bodies},
J. Math. Phys. {\bf 33} (1992), 1047-1053.
\bibitem{Ren94}  A.D.~Rendall, {\em The Newtonian limit for asymptotically flat solutions
of the Vlasov-Einstein system},  Comm. Math. Phys.  {\bf 163}  (1994),  89-112.
\bibitem{Scho86} S.~Schochet, {\em Symmetric hyperbolic systems with a large parameter},
Comm. partial differential equations, {\bf 11} (1986), 1627-1651.
\bibitem{Scho88}  S.~Schochet, {\em Asymptotics for symmetric hyperbolic systems with a
large parameter}, J. differential equations {\bf 75} (1988), 1-27.
\bibitem{Ukai86} S.~Ukai, {\em The incompressible limit and the initial layer
of the compressible Euler equations}, J. Math. Kyoto Univ. {\bf 26} (1986), 323-331.

\end{thebibliography}
